\documentclass[preprint,aps,12pt,preprintnumbers,eqsecnum,nofootinbib]{revtex4}
\usepackage{graphicx}
\usepackage{subfigure}

\usepackage{color}
\usepackage{amssymb,amsmath}
\usepackage{epstopdf}

\newcommand{\beq}{\begin{equation}}
\newcommand{\enq}{\end{equation}}

\begin{document}
%
%  V2 for the arXiv
% Title of paper
\title{\vspace*{0.5in} 
Dante's Waterfall
\vskip 0.1in}
\author{Christopher D. Carone}\email[]{cdcaro@wm.edu}
\author{Joshua Erlich}\email[]{jxerli@wm.edu}
\author{Anuraag Sensharma}\email[]{asensharma@email.wm.edu}
\author{Zhen Wang}\email[]{zwang01@email.wm.edu}

\affiliation{High Energy Theory Group, Department of Physics,
College of William and Mary, Williamsburg, VA 23187-8795}
\date{September dd, 2014}
\date{\today}

\begin{abstract}
We describe a hybrid axion-monodromy inflation model motivated by the Dante's Inferno scenario.  In Dante's Inferno, a  two-field potential features a stable trench along which a 
linear combination of the two fields slowly rolls, rendering the dynamics essentially identical to that of single-field chaotic inflation.  A shift symmetry allows for the Lyth bound to be 
effectively evaded as in other axion-monodromy models.  In our proposal, the potential is concave downward near the origin and the inflaton trajectory is a gradual downward spiral,
ending at a point where the trench becomes unstable.  There, the fields begin falling rapidly towards the minimum of the potential and inflation terminates as in a hybrid model.  We 
find parameter choices that reproduce observed features of the cosmic microwave background, and discuss our model in light of recent results from the BICEP2 and Planck experiments.    
\end{abstract}
\maketitle

\section{Introduction}
The  $B$-modes in the polarization of the cosmic microwave background (CMB) reported by the BICEP2 collaboration~\cite{Ade:2014xna}  may be due to primordial gravitational 
waves~\cite{Seljak:1996gy,Kamionkowski:1996zd}, or may be due to conventional polarization-dependent processes such as scattering off of galactic dust~\cite{Flauger:2014qra,Mortonson:2014bja}, as suggested by recent 
measurements by the Planck collaboration~\cite{Adam:2014oea}.  Tensor modes in primordial gravitational waves could produce an observably large $B$-mode polarization signal if the scale of inflation is high enough, typically 
around the GUT scale.  However,  the Lyth bound~\cite{Lyth:1996im} implies that generically in such scenarios, the inflaton   varies over a super-Planckian range of field values during inflation.  This would render an effective field theory 
treatment invalid, so possibilities for evading the Lyth bound are of practical interest.  One possibility is that the slow-roll parameter $\epsilon$ varies by a large multiplicative factor during inflation, which renders the Lyth-bound 
analysis invalid~\cite{Carrillo-Gonzalez:2014tia,Carone:2014lba}.  Another possibility is that the inflaton is an axion with an associated shift symmetry.  In such a scenario,  super-Planckian values of the inflaton field are identified with 
sub-Planckian values plus additional fluxes of one or more other fields~\cite{McAllister:2008hb}.  These axion-monodromy models provide a framework consistent with effective field theory which could accommodate an observably large amplitude in tensor modes.

A simplified scenario incorporating the axion-monodromy idea, improving on inflation models with two axions \cite{Kim:2004rp},  is known as Dante's Inferno~\cite{Berg:2009tg}.  The two axions of the Dante's Inferno model play different roles: one has an explicitly broken shift symmetry while the other maintains a discrete shift symmetry. The periodic nature of the two-field potential gives rise to a trench that extends down to the minimum of the potential. 
The inflaton field is identified with the linear combination of fields that slowly rolls down the trench, and can wind many times during inflation while neither of the two fields ever takes super-Planckian values.  Hence, this model is amenable to  an effective-field-theory treatment even if significant power in tensor modes is produced during inflation. The inflationary dynamics in the Dante's Inferno scenario is controlled by the shape of the potential along the one-dimensional trench, and the scenario makes the same predictions as a single-field chaotic inflation model.  
The Lagrangian for the two fields, $r$ and $\theta$, in the Dante's Inferno model is given by~\cite{Berg:2009tg}
\begin{equation}
{\cal L}=\frac{1}{2}(\partial_\mu r)^2+\frac12(\partial_\mu\theta)^2-V(r,\theta) \,\, , \label{eq:DI-Lag1}
\end{equation}
where the potential $V(r,\theta)$ respects the discrete shift symmetry in $\theta$ and the broken shift symmetry in $r$: \begin{equation}
V(r,\theta)=W(r)+\Lambda^4\left[1-\cos\left(\frac{r}{f_r}-\frac{\theta}{f_\theta}\right)\right].
\label{eq:DI-Lag2}
\end{equation}
\begin{figure}[t]
\includegraphics[width = 0.5\textwidth]{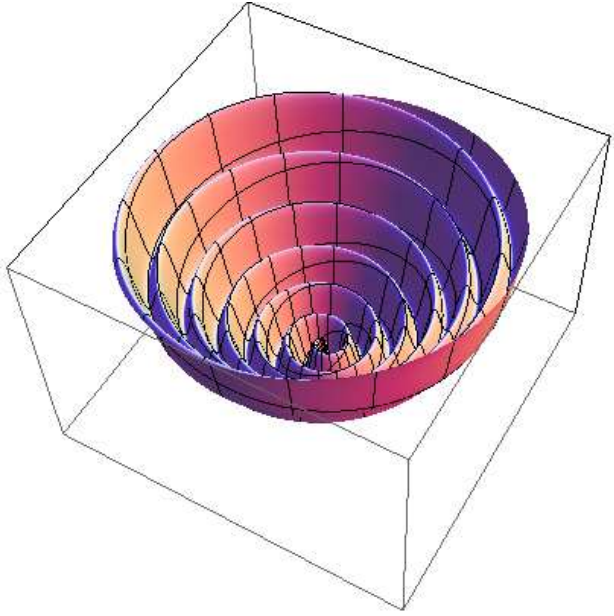}
\caption{The potential as a function of $r$ and $\theta$ in Dante's Inferno with a quadratic shift-symmetry-breaking potential, as in Ref.~\cite{Berg:2009tg}. The field $\theta$ is represented in cylindrical coordinates with 
period $2 \pi f_\theta$.}
\label{fig:Dante-trench}
\end{figure}
The potential $W(r)$ explicitly breaks the shift symmetry of the field $r$, which in a string theory realization could be due to nonperturbative effects related to moduli 
stabilization~\cite{Berg:2009tg}.   Assuming $W(r)=\frac{1}{2} m^2 r^2$, the cosine term in $V(r,\theta)$ gives rise to a staircase-like trench in the potential, as shown in 
Fig.~\ref{fig:Dante-trench}, where the coordinate $\theta$ is wrapped in cylindrical coordinates to reflect the shift symmetry.    With this choice of $W(r)$, the inflaton accelerates 
along the trench, both before and for some time after the end inflation, with the transition occurring when the slow roll conditions ({\em e.g.} $\epsilon < 1$) are violated.  The dynamics of the inflaton field can be described by an 
effective one-dimensional inflaton potential that is quadratic~\cite{Berg:2009tg}, so that the predictions for inflationary observables are identical to those of an analogous 
chaotic inflation model~\cite{Linde:1983gd}.  In particular, the scenario allows for relatively large power in tensor modes, with ratio of tensor to scalar amplitudes $r=0.14$.

We present a variation of the Dante's Inferno scenario in which the inflaton trench becomes unstable for a range of inflaton field values.  In this scenario, the slow-roll conditions break down only after the inflaton rolls 
off the trench and begins moving rapidly in an independent direction in field space.  Thus, inflation ends as in a hybrid model.  In hybrid inflation, the waterfall field has an effective squared
mass that depends on the inflaton field value.  At a critical point, this squared mass becomes negative and the system rapidly evolves to its global minimum.  In our scenario, the
same is true for a linear combination of the fields $r$ and $\theta$: one linear combination is identified as the inflaton and the effective squared mass of the remaining combination
depends on the inflaton field value.  When this squared mass becomes negative, the combination of fields that rolls quickly towards the potential minimum (and then oscillates about it)  
acts as the waterfall field of hybrid inflation~\cite{Linde:1993cn}.  Hence, we  refer to this scenario as Dante's Waterfall.  The model has the same Lagrangian as the Dante's Inferno model, 
Eqs.~(\ref{eq:DI-Lag1})-(\ref{eq:DI-Lag2}), but with a symmetry-breaking potential 
\begin{equation}
W(r)=-\frac{1}{2}m^2r^2+\frac{\lambda}{4!} r^4+\frac{3}{2}\frac{m^4}{\lambda}.
\label{eq:W}\end{equation}
\begin{figure}[t]
\includegraphics[width = 0.5\textwidth]{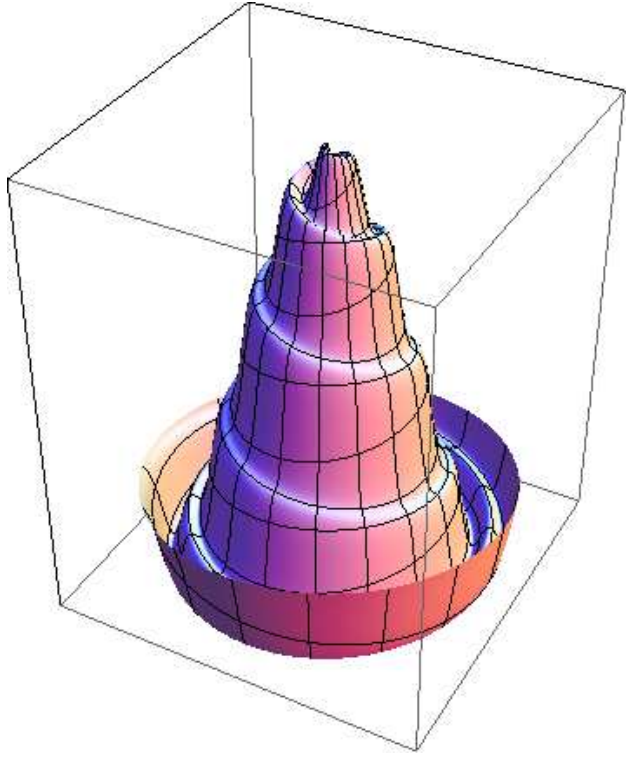}
\caption{The potential as a function of $r$ and $\theta$ in Dante's Waterfall, with symmetry-breaking potential $W(r)$ as in Eq.~(\ref{eq:W}). The field $\theta$ is represented in cylindrical coordinates with 
period $2 \pi f_\theta$.}
\label{fig:DW}
\end{figure}
An inflation model with a similar symmetry-breaking potential has been considered recently in Ref.~\cite{mcdonald}.
The last term in Eq.~(\ref{eq:W}) is included so that the full potential $V(r,\theta)$ vanishes at its global minimum.  This is the usual fine-tuning of the cosmological constant.
With this form for $W(r)$, the potential $V(r,\theta)$ is as in Fig.~\ref{fig:DW}.  In the typical Dante's Inferno scenario,  the trench is  unstable only for large field values not relevant 
during inflation.  However, in the present scenario, depending on the choice of model parameters, it is possible for the trench to become unstable for a range of intermediate field 
values.  This is the scenario we consider here.  We analyze cosmological observables analytically under certain assumptions in Sec.~\ref{sec:themod}, and more generally in 
Sec.~\ref{sec:numerics}.  We conclude in Sec.~\ref{sec:conc}.

\section{Single-field Effective Theory} \label{sec:themod}
By a field rotation the potential, Eqs.~(\ref{eq:DI-Lag2})-(\ref{eq:W}), can be written
\begin{equation}
V = -\frac{1}{2} m^2 r^2 + \frac{\lambda}{4!} \,r^4 + \frac{3}{2} \frac{m^4 }{\lambda} 
+ \Lambda^4 \left[1-\cos(\tilde{r}/f)\right] \,\,\, , \label{eq:fullpot}
\end{equation}
where $r=c\,\tilde{r}+s \, \tilde{\theta}$, $\theta=c\,\tilde{\theta}-s\,\tilde{r}$, and $s\equiv \sin\xi$, $c\equiv\cos \xi$ define the field rotation.  In terms of the parameters in
Eqs.~(\ref{eq:DI-Lag2})-(\ref{eq:W}),
\begin{equation}
\sin\xi = \frac{f_r}{\sqrt{f_r^2+f_\theta^2}}\, ,\,\,\,\,\, \cos\xi=\frac{f_\theta}{\sqrt{f_r^2+f_\theta^2}} \, ,\,\,\,\, \mbox{ and }  \,\,\,\,\, f=\frac{f_r f_\theta}{\sqrt{f_r^2+f_\theta^2}} \,.
\end{equation}
We assume for most of our analysis that $\sin\xi\ll1$,  or equivalently, $f_r\ll f_\theta$.  The trench in field space is given by $\partial V(\tilde{r},\tilde{\theta})/\partial\tilde{r}=0$, or 
\begin{equation}
-m^2 c\, r + \frac{\lambda}{6} c\, r^3 + \frac{\Lambda^4}{f} \sin(\tilde{r}/f) =0 \,\,\, .
\label{eq:trench}
\end{equation}
We have expressed this equation in terms of $r$ and $\tilde{r}=c\, r -s\, \theta$ to present it in a notationally compact form and to facilitate comparison with Ref.~\cite{Berg:2009tg}
where the same mixed notation was used.  The trench defines an effective one-dimensional potential along which the inflaton field slowly rolls.  During inflation, 
motion along the trench continues provided the stability condition $\partial^2 V(\tilde{r},\tilde{\theta})/\partial\tilde{r}^2>0$ is satisfied.  The end of inflation occurs at a point satisfying 
Eq.~(\ref{eq:trench}) where $\partial^2 V/\partial\tilde{r}^2=0$, that is
\begin{equation}
-m^2 c^2 + \frac{\lambda}{2} c^2 r^2 + \frac{\Lambda^4}{f^2} \cos(\tilde{r}/f)=0 \,\,\,.
\label{eq:end}
\end{equation}
As we will see, the fields then rapidly deviate from their original trajectory and approach the global minimum of the potential.

As in the model of Ref.~\cite{Berg:2009tg}, there are limits of our model where inflation can be described by the evolution of a single field with canonically normalized kinetic terms.   
This is the field associated with the direction in field space given by Eq.~(\ref{eq:trench}), assuming one is in a field region where  
\begin{equation}
|cf (m^2 r - \lambda  r^3/6)|/\Lambda^4 \ll 1\,\,\, ,
\label{eq:condB}
\end{equation}
and one chooses
\begin{equation}
s\, c\, f^2 m^2 / \Lambda^4 \ll 1 \,\,\, .
\label{eq:condA}
\end{equation}
Eq.~(\ref{eq:condB}) allows one to approximate $\sin(\tilde{r}/f) \approx \tilde{r}/f$, leading to a linear relationship
between $\tilde{r}$ and $\tilde{\theta}$:
\begin{equation}
\tilde{r} = \left[\frac{ f^2 m^2 s \, c}{\Lambda^4 - f^2 m^2 c^2} \right] \tilde{\theta} \approx s \, c \,\left(\frac{f^2 m^2}{\Lambda^4} \right) \tilde{\theta}  \,\,\, .
\label{eq:roft}
\end{equation}
Identifying $\tilde{\theta} \equiv \phi$ as the inflaton in the single field effective theory,  Eq.~(\ref{eq:condA}) assures that the inflaton kinetic terms are canonical, up to small corrections of 
order $(s \,c\, m^2 f^2 / \Lambda^4)^2$.   We will make the further simplifying assumption in what follows that both $s \ll 1$ and $f^2 m^2 / \Lambda^4 \ll 1$.

Eliminating $\tilde{r}(\tilde{\theta})$ from Eq.~(\ref{eq:fullpot}) using Eq.~(\ref{eq:roft}), one obtains the effective single-field inflaton potential
\begin{equation}
V_{\rm eff} = -\frac{1}{2} \, m_{\rm eff}^2 \, \phi^2 + V_0 \,\,\,,
\label{eq:veff}
\end{equation}
where
\begin{equation}
m_{\rm eff} \equiv  m \, s  \, , \,\,\,\, V_0 \equiv \frac{3}{2 \lambda} \, m^4 \, , \,\,\,\,\, \mbox{and} \,\,\,\,\,  \phi \equiv \tilde{\theta}.
\end{equation}
In the case where $f_r \ll f_\theta$,   $s \approx f_r/ f_\theta$ and $m_{\rm eff} = m f_r / f_\theta$, as in the model of Ref.~\cite{Berg:2009tg}.

We now use this effective description to study a point in model parameter space that is viable.  One should keep in mind that such solutions are
approximate since the assumptions that justify the single-field approximation will generally fail somewhere near the end of the trajectory in field space, the point where
the waterfall occurs, as determined by Eq.~(\ref{eq:end}).  This affects the calculation of the number of $e$-folds of inflation occurring between the initial 
and final field values, $\phi_i$ and $\phi_f$, respectively, which we aim to hold fixed between $50$ and $60$.  However, since most of inflation occurs on the 
part of the trajectory where the single-field approximation is valid, our solutions should be qualitatively trustworthy, as we check in Sec.~\ref{sec:numerics}.   This is not very different 
from the case in non-hybrid inflation models, where one computes the number of $e$-folds by first declaring that the end of inflation corresponds to the value of the slow-roll 
parameter $\epsilon=1$.  Here, we define the end of inflation as $\phi_f  = \tilde{\theta}_f$, where $(\tilde{r}_f, \tilde{\theta}_f$) lies on a trench and 
satisfies $\partial^2 V/\partial\tilde{r}^2=0$.  

To find an acceptable solution, we first require that our effective theory provide the correct values for the spectral index $n_s$ and the the amplitude of the scalar perturbations 
$\Delta_R^2$, since these quantities are relatively well measured.  For definiteness, we assume the experimental
central values~\cite{Ade:2013uln}.   We first define the slow-roll parameters
\beq
\epsilon \equiv \frac{M_{P}^2}{16 \pi}\left(\frac{V'}{V}\right)^2, \quad
\eta \equiv \frac{M_{P}^2}{8 \pi} \frac{V''}{V}, \quad
\gamma \equiv \frac{M_{P}^4}{64 \pi^2} \frac{V' V'''}{V^2} \,\,\, ,
\enq
where the derivatives of the potential are with respect to $\phi$. In general, it follows from Eq.~(\ref{eq:veff}) that $\gamma=0$,  
\begin{equation}
\epsilon = \frac{M_P^2}{4 \pi} \frac{\phi^2}{\left(2 V_0/m_{\rm eff}^2-\phi^2\right)^2} \,\,\, 
\,\,\,\, \mbox{ and }  \,\,\,\,\,
\eta = - \frac{M_P^2}{4 \pi} \frac{1}{\left(2 V_0/m_{\rm eff}^2-\phi^2\right)} \,\,\, .
\label{eq:epseta}\end{equation}
The spectral index $n_s$ and scalar amplitude $\Delta_R^2$ may be expressed as
\begin{equation}
n_s= \left[1-6\epsilon+2\eta\right]_{\phi=\phi_i},
\label{eq:ns}
\end{equation}
\begin{equation}
\Delta_R^2=\left[\frac{8}{3M_P^4}\frac{V}{\epsilon}\right]_{\phi=\phi_i}
\label{eq:DeltaRsq}
\end{equation}
where $\phi_i$ is the value of the inflaton field  50-60 $e$-folds before the end of inflation, when the largest distance scales that are currently observable exited the horizon.
Using Eq.~(\ref{eq:epseta}) one finds
\begin{equation}
n_s = 1 -\frac{M_P^2}{4\pi} \left[ \frac{6 \phi_i^2}{(2 V_0/m_{\rm eff}^2-\phi_i^2)^2} + \frac{2}{(2 V_0/m_{\rm eff}^2-\phi_i^2)}\right]
 \,\,\, ,\label{eq:nseq}
\end{equation}
\begin{equation}
\Delta_R^2 = \frac{16 \pi}{3 M_P^6} \frac{m_{\rm eff}^2}{\phi_i^2} \left[\frac{2 V_0}{m_{\rm eff}^2} - \phi_i^2 \right]^3 \,\,\, . 
\label{eq:dreq}
\end{equation}
Our formulae assume $2 V_0/m_{\rm eff}^2-\phi_i^2>0$, which is consistent with our numerical results.   We work henceforth in units where $M_P=1$.

Taking $m_{\rm eff}$ as an input parameter, and setting $n_s=0.9603$ and $\Delta_R^2=2.2 \times 10^{-9}$~\cite{Ade:2013uln}, we find that Eqs.~(\ref{eq:nseq}) 
and (\ref{eq:dreq}) only have solutions if $m_{\rm eff} \alt 8.31 \times 10^{-7}$.    For example, the choice $m_{\rm eff} = 1.2 \times 10^{-7}$
yields
\begin{equation}
V_0 = 2.885 \times 10^{-14} \,\,\,\,\, \mbox{ and } \,\,\,\,\, \phi_i = 0.0838 \,\,.
\label{eq:mvp1}
\end{equation}
We can now ask whether there is an acceptable trajectory in the full theory with $\tilde{\theta}$ beginning at $\phi_i$, and terminating
at a point where $d^2V / d\tilde{r}^2=0$ such that $50$ to $60$ $e$-folds of inflation is obtained.  With $m_{\rm eff}$ and $V_0$ fixed, 
we constrain two degrees of freedom in the parameter space of the complete theory.   We choose the value of $s$ and fix
\begin{equation}
m = m_{\rm eff}/s 
\label{eq:mfix}
\end{equation}
and
\begin{equation}
\lambda = \frac{3}{2} \frac{m_{\rm eff}^4}{s^4} \frac{1}{V_0} \,\,\, .
\label{eq:lamfix}
\end{equation}
Specifying $\Lambda$ and $f$ then completely determines Eq.~(\ref{eq:fullpot}).   Consider the following choice of
parameters, that are consistent with Eqs.~(\ref{eq:mvp1}), (\ref{eq:mfix}) and (\ref{eq:lamfix}):
\begin{eqnarray}
s &=& 0.0010 \,\, ,\nonumber \\
\lambda &=& 1.078 \times 10^{-2} \,\, ,\nonumber \\
\Lambda &=& 0.0001 \,\, ,\nonumber \\
m &=& 0.00012 \,\, , \nonumber \\
f & = & 2.453 \times 10^{-5} \,\, .
\label{eq:parameters}
\end{eqnarray}
One can verify that the following points in field space are continuously connected by a solution to Eq.~(\ref{eq:trench})
\begin{eqnarray}
(\tilde{r} , \tilde{\theta})_i &=& (8.099 \times 10^{-6} , \,\,  8.377 \times 10^{-2} ) \,\, ,\nonumber \\
(\tilde{r}, \tilde{\theta})_f &=& (3.647 \times 10^{-5} ,  \,\, 2.485 \times 10^{-1})   \,\, .  
\label{eq:startend}
\end{eqnarray}
In addition, $(\tilde{r}, \tilde{\theta})_f$ satisfies Eq.~(\ref{eq:end}).  Identifying $\phi_f=\tilde{\theta}_f$, one can now evaluate the
number of $e$-folds,
\begin{eqnarray}
N&=&\frac{2\sqrt{\pi}}{M_P}\int_{\phi_i}^{\phi_f}\frac{1}{\sqrt{\epsilon}}\,d\phi
\label{eq:N} \\
&=&\frac{4 \pi}{M_P^2} \left[ \frac{2 V_0}{m_{\rm eff}^2} \ln (\phi_f/\phi_i) - \frac{1}{2} (\phi_f^2 - \phi_i^2) \right]  
\,\,\,, \label{eq:neq}
\end{eqnarray}
from which one obtains $N=54.4$.   

The remaining cosmological parameters of interest can be expressed in terms of the slow-roll parameters.  We represent
the ratio of tensor-to-scalar amplitudes by $\underline{r}$ (to distinguish it from the field $r$), which is given by
\begin{equation}
\underline{r} = \left[16 \, \epsilon \right]_{\phi=\phi_i}
\label{eq:r},
\end{equation}
and the running of the spectral index by 
\begin{equation}
n_r = \left[16 \epsilon \eta - 24 \epsilon^2-2\gamma \right]_{\phi=\phi_i} \,\,\, . 
\label{eq:nr}
\end{equation}   
In the present example, one finds
\begin{eqnarray}
\underline{r}     &=& 5.585 \times 10^{-4}\,\, ,\nonumber \\
n_r &=& -1.114 \times 10^{-5}  \,\, .
\end{eqnarray}
There are consistent with current bounds~\cite{Ade:2013uln}, given the lingering questions surrounding the current BICEP2 measurement 
of $\underline{r}$. We will discuss larger possible values of $\underline{r}$ later in this section.

It is worth returning to the inequalities that we assumed at the beginning of this section.  For the example given,
the condition in Eq.~(\ref{eq:condA}) is satisfied, with the left-hand-side evaluating to $\approx 8.7 \times 10^{-5}$.   The condition
in Eq.~(\ref{eq:condB}) is satisfied at the beginning of the trajectory, where the left-hand-side is $\approx 0.324$,
and fails at the end, as we anticipated earlier, where the same quantity is $\approx 0.996$.   

\begin{figure}[h]
\includegraphics[width=2.5in]{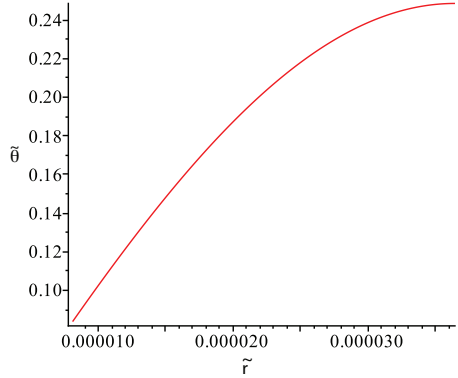}
\centering
\caption{Trajectory in field space, $\tilde{\theta}(\tilde{r})$, during inflation.} \label{fig:one}
\end{figure}

To better visualize the solution, we first note that in the original $(r,\theta)$ coordinate system, the global minimum
is located at
\begin{equation}
r_{min}= \sqrt{\frac{6}{\lambda}} m = 2.831 \times 10^{-3}\,\,\, ,
\end{equation}
while the initial and final $r$ values are
\begin{eqnarray}
r_i &=& 9.187 \times 10^{-5} \nonumber \\
r_f &=& 2.850 \times 10^{-4}    \,\,\, .
\end{eqnarray}
The trajectory in this example is far from the global minimum at positive $r$  and moving toward it, as one might expect.  A plot 
of the trajectory in $\tilde{r}-\tilde{\theta}$ space during inflation is shown in Fig.~\ref{fig:one}.

\begin{figure}[h]
\subfigure{\includegraphics[width = 2.8in]{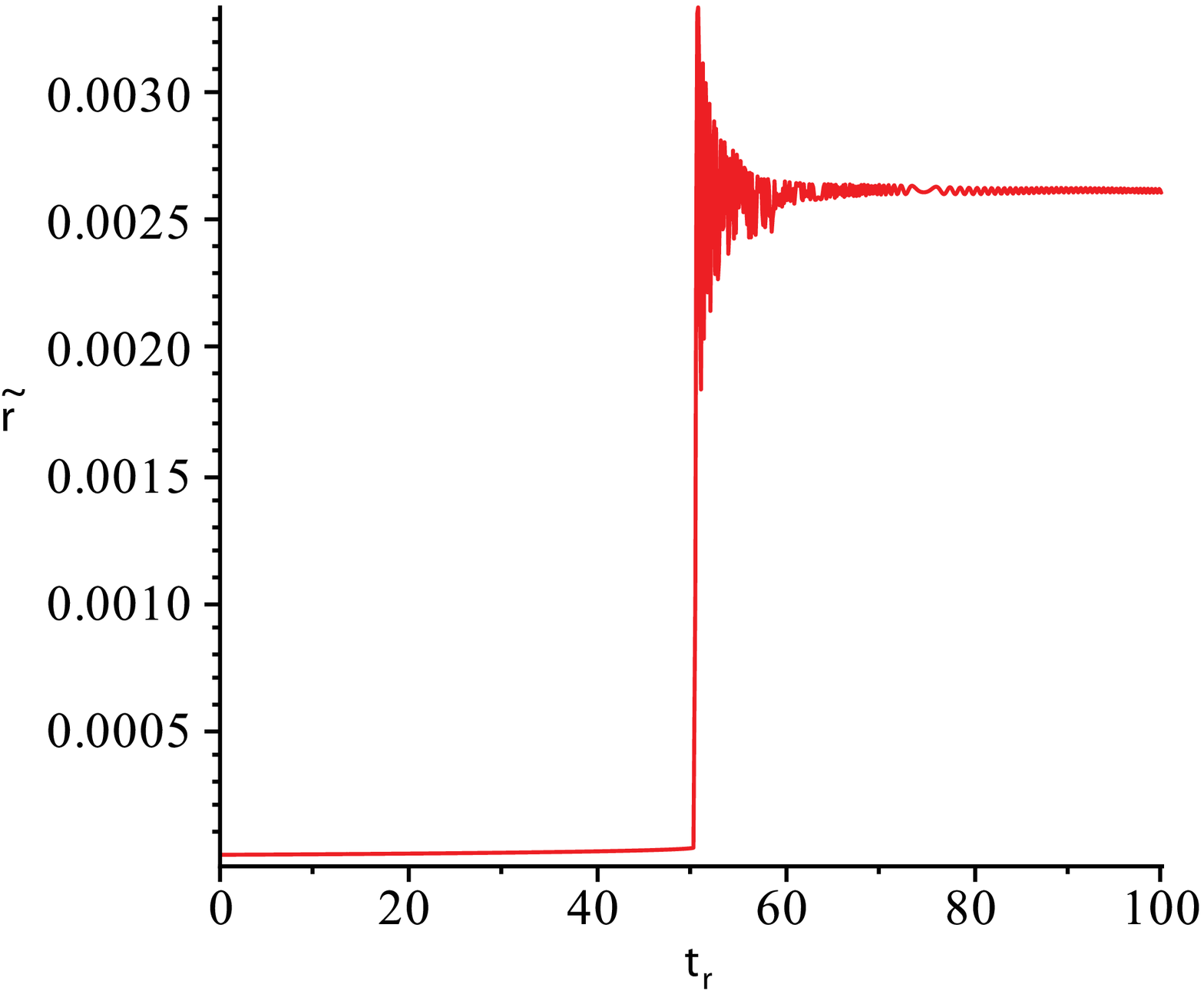}}
\subfigure{\includegraphics[width = 2.5in]{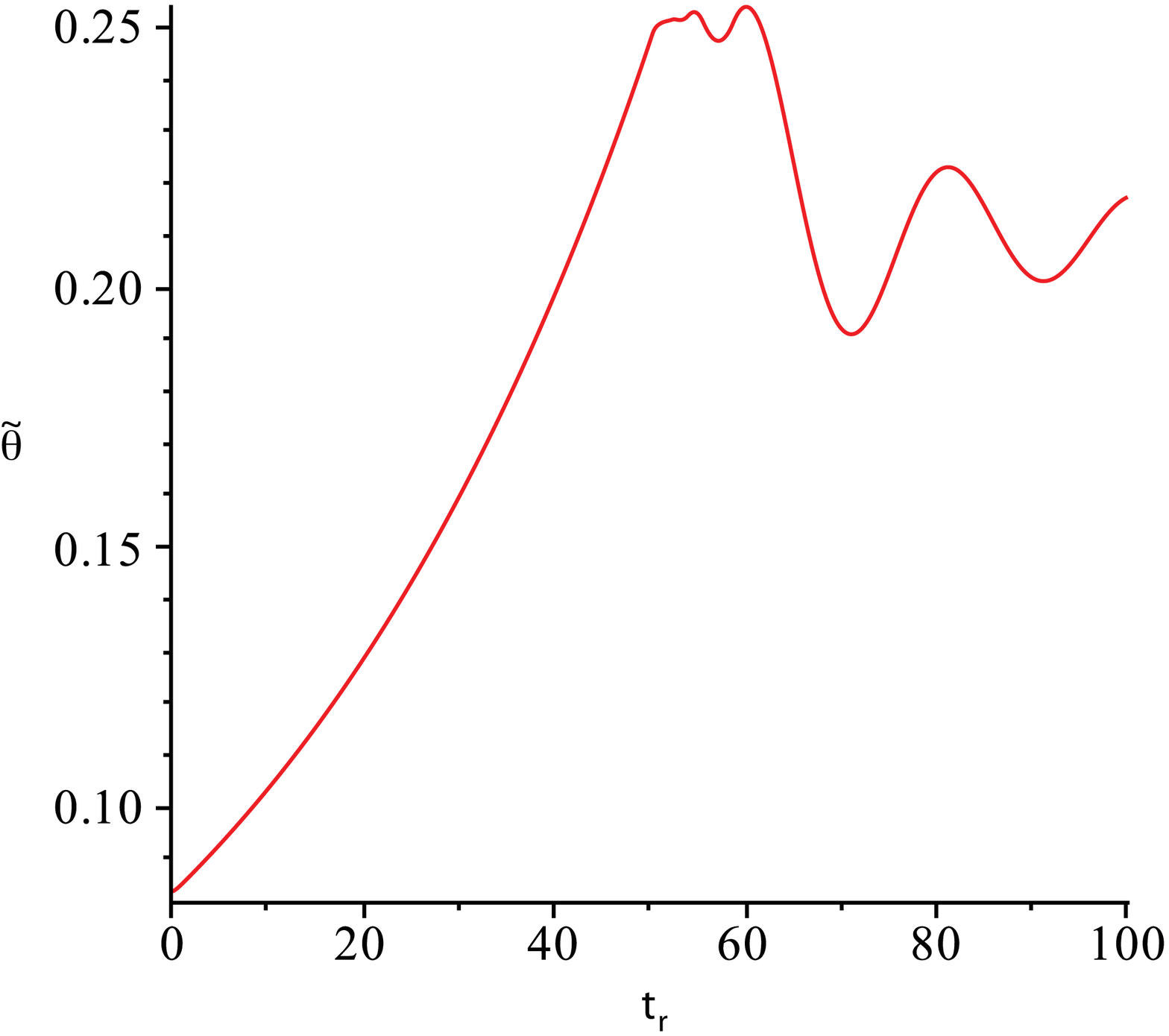}}
\centering
\caption{Time-evolution of the fields: (a) $\tilde{r}(t)$, (b) $\tilde{\theta}(t)$. The time is in units of the
inverse Hubble parameter at the beginning of inflation and the Planck mass has been set to one.} \label{fig:two}
\end{figure}

One can confirm the end of inflation in this example by studying the time evolution of the fields in the full theory, $\tilde{r}(t)$
and $\tilde{\theta}(t)$, which satisfy the coupled equations of motion
\begin{eqnarray}
\ddot{\tilde{r}} + 3 H \dot{\tilde{r}} + \frac{\partial V}{\partial \tilde{r}}&=&0  \,\,\, ,\nonumber \\
\ddot{\tilde{\theta}} + 3 H \dot{\tilde{\theta}} + \frac{\partial V}{\partial \tilde{\theta}}&=&0 \,\,\, .
\label{eq:eom}
\end{eqnarray}
For definiteness, we assume that $\dot{\tilde{r}}$ and $\dot{\tilde{\theta}}$ are initially vanishing; qualitatively similar
solutions are obtained for other choices, providing that the slow-roll conditions are satisfied.
The results are shown in Fig.~\ref{fig:two}, with the time variable $t_r=H_0 \, t$ 
where $H_0 \equiv H(t=0)$ is the Hubble parameter at the beginning of inflation.
Notice that $\tilde{\theta}(t_r)$ pauses for a brief interval near $\tilde{\theta}_f$ (around $t_r\approx 50$), at precisely the 
same time that  $\tilde{r}(t_r)$ rapidly increases away from $\tilde{r}_f$:  this is the waterfall.   The fields then oscillate as they approach 
the global minimum, the period when reheating presumably occurs.

The example we have presented is useful in illustrating the qualitative features of a typical solution.  We now investigate whether
solutions with larger values of the parameter $\underline{r}$, {\em i.e.} larger tensor perturbations, are possible.  Given the constraints of
Eqs.~(\ref{eq:nseq}) and (\ref{eq:dreq}), specification of $m_{\rm eff}$ determines $\phi_i$ and hence also the parameter $\underline{r}$ in
our effective theory.   It follows that
\begin{equation}
\underline{r}(m_{\rm eff})= \frac{2}{9 \pi C_0^2} \left[ C_1 \pm (C_1^2 - 4 \, C_0 \, m_{\rm eff}^2)^{1/2} \right]  \,\,\, ,
\label{eq:romeff}
\end{equation}
where
\begin{equation}
C_0 = \left[\Delta_R^2/(144 \,\pi) \right]^{1/3}  \,\,\,\,\, \mbox{ and } \,\,\,\,\, C_1 = 6\, \pi \,C_0^2 \, (1-n_s) \,\,\,.
\end{equation}
Numerically, $C_0=1.694 \times 10^{-4}$ and $C_1 = 2.164 \times 10^{-8}$.   For these values, Eq.~(\ref{eq:romeff}) is 
maximized when $\underline{r}_{\rm max+} = 0.107$ or $\underline{r}_{\rm max-}=0.053$, depending on the sign of the square root, which
corresponds to different possible solutions for $\phi_i$.   We can  make further progress by considering the number of $e$-folds, 
given in Eq.~(\ref{eq:neq}).  As a function of $\phi_f$, this expression is maximized when $\phi_f^2 = 2 V_0/m_{\rm eff}^2$.   The 
value at the maximum, $N_{\rm max}$, is thus a function of $m_{\rm eff}$, like $\underline{r}$, and depends on the same sign choice appearing in Eq.~(\ref{eq:romeff}).   
We find that for the positive square root, $N_{\rm max}$ is below $42.4$ for any $m_{\rm eff}$; hence, these solutions are excluded.   For the 
negative square root, $N_{\rm max}$ falls below the desired range, $50$ to $60$, before $m_{\rm eff}$ is large enough to yield $\underline{r}_{\rm max-}=0.053$.  
We find numerically that $N>50$ forces $\underline{r} < 0.03$.   Hence, we expect on general grounds that
\begin{equation}
\underline{r} < 0.03 \,\,\, ,
\end{equation}
provided that Eq.~(\ref{eq:veff}) is an accurate effective description of the theory.  Whether a choice of parameters and field trajectory exists in the complete 
theory for which this bound is saturated is not guaranteed.  However, it is not hard to discover solutions that are of order this bound.  Following the approach of this
section, one can check, for example, that the parameter choice
\begin{eqnarray}
s &=& 0.01 \,\, ,\nonumber \\
\lambda &=& 1.635 \times 10^{-5} \,\, ,\nonumber \\
\Lambda &=& 5.0 \times 10^{-5} \,\, ,\nonumber \\
m &=& 5.0 \times 10^{-5} \,\, , \nonumber \\
f & = & 2.610 \times 10^{-7} \,\, ,
\end{eqnarray}
is consistent with the trajectory
\begin{eqnarray}
(\tilde{r} , \tilde{\theta})_i &=& (1.120 \times 10^{-7} , \,\,  0.406 ) \,\, ,\nonumber \\
(\tilde{r}, \tilde{\theta})_f &=& (4.099 \times 10^{-7} ,  \,\,  1.105 )   \,\, .  
\end{eqnarray}
This leads to the values $\underline{r}=0.011$ and $N=51.1$

\section{Numerical Analysis}\label{sec:numerics}
In the previous section we obtained an approximation for the shape of the one-dimensional potential, Eq.~(\ref{eq:veff}), which
followed from the linear relation in Eq.~(\ref{eq:roft}).   This relation breaks down before the end of inflation.  In this section, we find the 
shape of the trench and compute observables numerically, allowing us to test the validity of our previous approximation. 

We again choose $f_r\ll f_\theta$ and identify $\tilde{\theta}$ as the inflaton field.  Along the trench, Eq.~(\ref{eq:trench}),  $\tilde{r}$ is non-dynamical 
to lowest order in $f_r/f_\theta$ and corrections to the $\tilde{\theta}$ kinetic terms are negligible.  This can be verified by differentiating  
Eq.~(\ref{eq:trench}), which yields
\beq
\frac{\dot{\tilde{r}}}{\dot{\tilde{\theta}}}=\frac{sc\, f^2 (m^2-\frac{1}{2}\lambda r^2)}{\Lambda^4
\cos(\tilde{r}/f)-c^2\, f^2(m^2-\frac{1}{2}\lambda r^2) } \,\, .
\label{eq:kinetic}
\enq
In the region of field space where $m^2>\frac12\lambda r^2$, as long as
\beq
\frac{\Lambda^4\cos(\tilde{r}/f)}{f^2}\geq c^2 \,(2m^2-\lambda r^2) \,\, ,
\enq
the kinetic terms for $\tilde{r}$ and $\tilde{\theta}$ sum to
\beq
\frac{1}{2}\dot{\tilde{r}}^2+\frac{1}{2}\dot{\tilde{\theta}}^2\leq(1+\tan^2{\xi})\frac{1}{2}\dot{\tilde{\theta}}^2 \,\, .
\enq
In this case, the $\tilde{\theta}$ kinetic terms remain canonically normalized to leading order in $f_r/f_\theta$.  
The potential of the effective single-field description of the theory is given by
\beq
V(\tilde{\theta}) \equiv V(\tilde{r}_t(\tilde{\theta}), \tilde{\theta}) \,\, ,
\label{eq:vsub}
\enq
where $\tilde{r}_t(\tilde{\theta})$ is the solution to the trench equation Eq.~(\ref{eq:trench}). Derivatives of Eq.~(\ref{eq:vsub})
with respect to $\tilde{\theta}$ can be computed numerically to obtain the slow-roll parameters and the inflationary
observables discussed in Sec.~\ref{sec:themod}.

To test the accuracy of the quadratic form of the effective single-field potential, Eq.~(\ref{eq:veff}), we evaluate observables following from
Eq.~(\ref{eq:vsub}) using the same parameters, Eqs.~(\ref{eq:parameters}) and (\ref{eq:startend}).  Following from Eqs.~(\ref{eq:ns}), (\ref{eq:DeltaRsq}), (\ref{eq:r}) 
and (\ref{eq:nr}),  we find that $( n_s, \Delta_R^2, \underline{r}, n_r)=(0.956, 1.833 \times 10^{-9},6.70 \times 10^{-4}, -1.47\times 10^{-5})$. 
The number of $e$-folds is determined by Eq.~(\ref{eq:N}), from which we obtain $N=49.44$, somewhat smaller than the value $N=54.4$ 
that followed from the approximations of Sec.~\ref{sec:themod}.   This exercise confirms that the approximation scheme of Sec.~\ref{sec:themod}
provides a qualitatively accurate solution for the set of cosmological quantities of interest: the breakdown in this scheme 
occurs close enough to the end of the inflationary trajectory that it does not substantially alter the qualitative results.

\begin{figure}[ht]
\includegraphics[width = 0.5\textwidth]{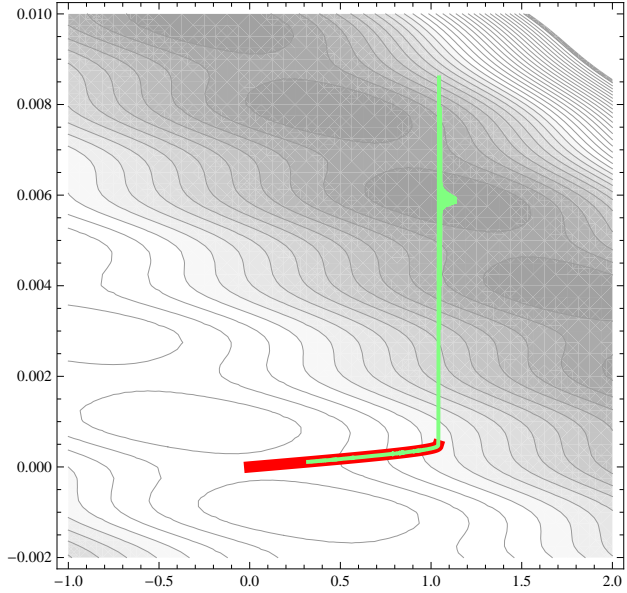}
\caption{Contour plot of the potential in terms of $\tilde{r}$ (vertical axis) and $\tilde{\theta}$ (horizontal axis). 
The thick, solid red line indicates the bottom of the trench. The inflationary trajectory is shown by the thin green line.}
\label{fig:potential}
\end{figure}

\begin{figure}[ht]
\subfigure{\includegraphics[width = 0.495\textwidth]{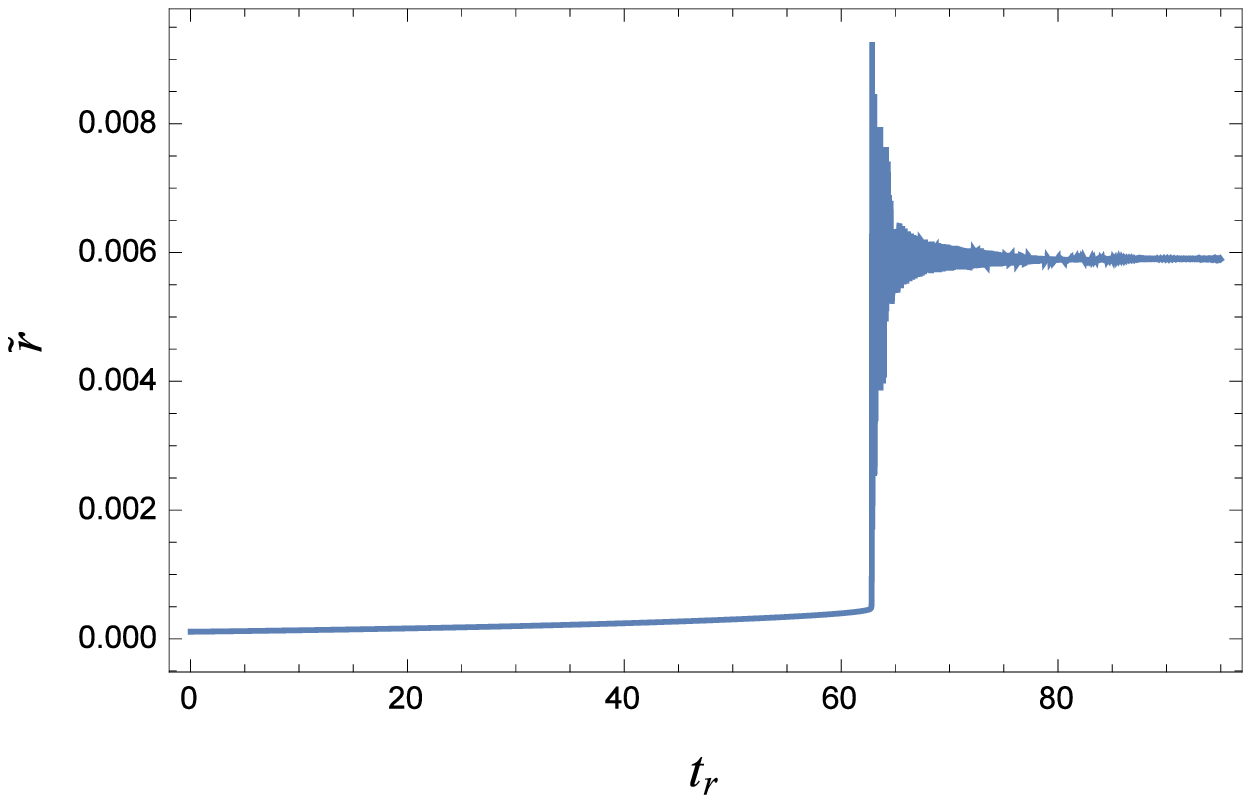}}
\subfigure{\includegraphics[width = 0.495\textwidth]{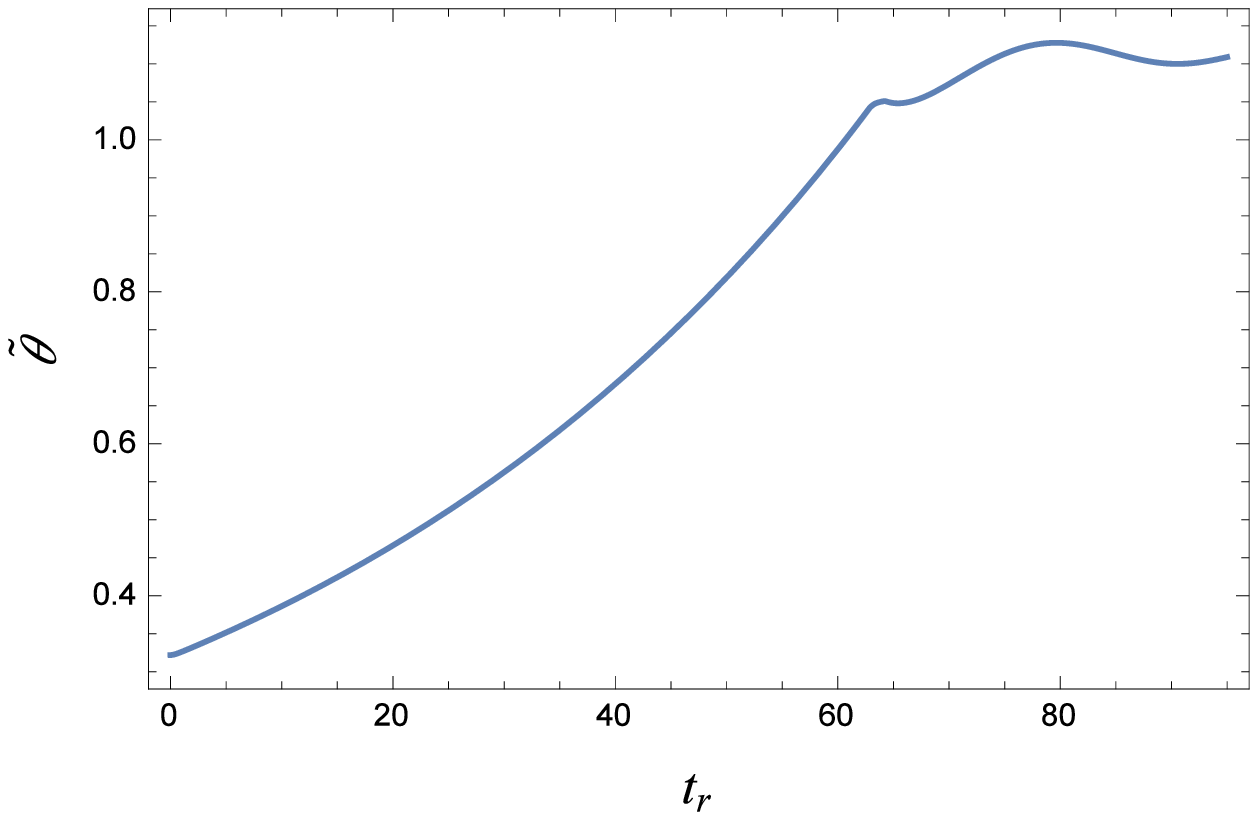}}
\caption{Dynamic solutions. The left graph shows $\tilde{r}(t_r)$ and the right graph shows $\tilde{\theta}(t_r)$. The time variable $t_r=H_0 t$ is scaled in units of Hubble time at the beginning of inflation.}
\label{fig:rttime}
\end{figure}

In the current numerical treatment, however, we can now find solutions that more exactly match the cosmological observables.  For example, 
with $(f/s, f/c, m ,\lambda, \Lambda)=(0.1043, 3.127 \times 10^{-4}, 1.367 \times 10^{-4}, 1.314\times 10^{-3}, 3.654 \times 10^{-4})$,  
$(\tilde{r}_i, \tilde{\theta}_i)=(1.112 \times 10^{-4}, 0.322)$ and $(\tilde{r}_f, \tilde{\theta}_f)=(4.738 \times 10^{-4}, 1.039)$, we obtain
\begin{eqnarray}
n_s&=&0.960 \, , \nonumber \\
\Delta^2_R&=&2.23 \times 10^{-9} \, , \nonumber \\
\underline{r} &=&7.45 \times 10^{-3} \, ,\nonumber \\
n_r&=&-1.42 \times 10^{-4} \, ,\nonumber \\
N&=&59.7 \,\, .
\label{eq:numericfit}
\end{eqnarray}
As with our previous solution, we may solve the coupled equations of motion for $\tilde{r}(t)$ and $\tilde{\theta}(t)$, with $\tilde{r}(0)=\tilde{r}_i$
and $\tilde{\theta}(0)=\tilde{\theta}_i$.  For definiteness, we again assume that the first time derivatives of the fields vanish at $t=0$, as discussed after Eq.~(\ref{eq:eom}). 
The trajectory in field space is shown in Fig.~\ref{fig:potential}, while $\tilde{r}(t)$ and $\tilde{\theta}(t)$ are shown in Fig.~\ref{fig:rttime}.
We can see that the system rolls along the trench until the instability is reached where inflation ends.   The system then moves quickly towards the 
global minimum of the potential.   We have checked that $\dot{\tilde{r}}(t)^2/\dot{\tilde{\theta}}(t)^2$ remains small along the portion of this
trajectory where inflation occurs, never exceeding $10^{-7}$, so that the classical wavefunction renormalization of the inflaton field is negligible.

\section{Conclusions} \label{sec:conc}

We have studied a new realization of hybrid inflation in a variant of an axion monodromy model known as Dante's Inferno~\cite{Berg:2009tg}.  By altering the 
assumed form of the shift-symmetry-breaking potential of one of the axion fields, the scalar potential in our model takes the form of
a Mexican hat with an indentation, or trench, spiraling down from its peak.  Inflation corresponds to slowly rolling down this trench until 
a point where the trench becomes shallow and can no longer support the motion; the system then evolves rapidly in the radial direction towards the global 
minimum of the potential.  After formulating an appropriate single-field approximation for the period of inflation,  we studied viable points in 
model parameter space where the amplitude of scalar perturbations, the spectral index, the running of the spectral index, and the number of $e$-folds of 
inflation are consistent with observational data.  In an approximation where the single-field potential could be studied analytically, we argued that, given the assumed form of the potential, the parameter $\underline{r}$, which reflects that power in tensor modes, could be no larger than $0.03$, and we found explicit solutions where the value 
was $\sim 0.01$.  Future measurements of the microwave background polarization, that may probe $\underline{r}>0.007$~\cite{Lazear:2014bga}, have the potential of detecting a gravity-wave signal of this size; observational results closer to those of BICEP2~\cite{Ade:2014xna} would exclude the model.  It would be interesting to consider in more detail the various possibilities for the post-inflationary 
dynamics and reheating in this scenario.  

%%%%%%%%%%%%%%%%%%%%%%%%%%%%%%%%%%%%%%%%%%%%%%%%%%%%%%%%%%%
\begin{acknowledgments}  
This work was supported by the NSF under Grant PHY-1068008.  A.S. thanks the William \& Mary Research
Experiences for Undergraduates (REU) program for it's support via NSF Grant PHY-1359364.  We thank Jackson Olsen for a careful reading of the manuscript.
\end{acknowledgments}
%%%%%%%%%%%%%%%%%%%%%%%%%%%%%%%%%%%%%%%%%%%%%%%%%%%%%%%%%%%     

{\em Note Added.} $-$ While our manuscript was under review, analysis of a similar model appeared in Ref.~\cite{Barenboim:2014vea}.  We were also made aware of another
hybrid monodromy model which appears in Ref.~\cite{Hertzberg:2014sza}.
                                                                                                                                                              
%\appendix
%\section{}

\end{document}